\pdfoutput=1

\documentclass[11pt,twoside,a4paper,cmspaper,final,collab]{cms-tdr}
\def\svnVersion{436162P}\def\cmsCernNoTag{CERN-EP-2017-125}\def\cmsCernDate{\today}\def\cmsMessage{Published in Physics Letters B as \href{http://dx.doi.org/10.1016/j.physletb.2017.11.001}{\doi{10.1016/j.physletb.2017.11.001.}}}

\begin{document}\cmsNoteHeader{EXO-16-026}

\hyphenation{had-ron-i-za-tion}
\hyphenation{cal-or-i-me-ter}
\hyphenation{de-vices}

\RCS$Revision: 429536 $
\RCS$x: svn+ssh://svn.cern.ch/reps/tdr2/notes/EXO-12-046/trunk/EXO-12-046.tex $
\RCS$Id: EXO-16-026.tex 429536 2017-10-14 13:32:47Z alverson $
\newlength\cmsFigWidth
\ifthenelse{\boolean{cms@external}}{\setlength\cmsFigWidth{0.83\columnwidth}}{\setlength\cmsFigWidth{0.43\textwidth}}
\ifthenelse{\boolean{cms@external}}{\providecommand{\cmsLeft}{top}}{\providecommand{\cmsLeft}{left}}
\ifthenelse{\boolean{cms@external}}{\providecommand{\cmsRight}{bottom}}{\providecommand{\cmsRight}{right}}
\newcommand{\x}{\ensuremath{\phantom{0}}}

\cmsNoteHeader{EXO-12-046}
\title{
Search for a heavy composite Majorana neutrino in the final state with two leptons and two quarks at $\sqrt{s} = 13\TeV $
}

\date{\today}

\abstract{
A search for physics beyond the standard model in the final state with two same-flavour leptons (electrons or muons) and two quarks produced in proton-proton collisions at $\sqrt{s} = 13\TeV$ is presented. The data were recorded by the CMS experiment at the CERN LHC and correspond to an integrated luminosity of 2.3\fbinv. The observed data are in good agreement with the standard model background prediction. The results of the measurement are interpreted in the framework of a recently proposed model in which a heavy Majorana neutrino, $\textrm{N}_{\ell}$, stems from a composite-fermion scenario. Exclusion limits are set for the first time on the mass of the heavy composite Majorana neutrino, $m_{\textrm{N}_{\ell}}$, and the compositeness scale $\Lambda$. For the case $m_{\textrm{N}_{\ell}} = \Lambda$, the existence of $\textrm{N}_{\Pe}$ ($\textrm{N}_{\mu}$) is excluded for masses up to 4.60 (4.70)\TeV at 95\% confidence level.
}
\hypersetup{%
pdfauthor={CMS Collaboration},%
pdftitle={Search for a heavy composite Majorana neutrino decaying to two leptons and two quarks at sqrt(s) = 13 TeV},
pdfsubject={CMS},%
pdfkeywords={CMS, Exotica}
}

\maketitle
\section{Introduction}
\label{Introduction}

Experimental evidence has promoted the standard model (SM) to the role of the reference theory for high-energy particle physics.
Despite its successes, there are several fundamental aspects of observed particle physics that lack a complete explanation within the SM.
One of these is the mass hierarchy of fermions, for which a possible solution
has been offered by composite-fermion models~\cite{Pati:1975md,harari,Greenberg:1974qb}.

In the composite-fermion scenario, quarks and leptons are assumed to have an internal substructure that should manifest itself at some sufficiently high energy scale, the compositeness scale $\Lambda$. Ordinary fermions are considered as bound states of some not yet observed fundamental constituents generically referred to as \emph{preons}.
Two model-independent properties~\cite{Terazawa,Eichten,Cabibbo,Baur} are experimentally relevant:
the existence of a contact interaction, in addition to the gauge interaction,
which represents an effective approach for describing the effects of the unknown internal dynamics of compositeness, and
the existence of excited states of quarks and leptons with masses lower than or equal to $\Lambda$.
A particular case of such excited states could be a heavy composite Majorana neutrino (HCMN), $\textrm{N}_{\ell}$ $(\ell = \Pe,\,\mu,\,\tau)$~\cite{O3,O2,O1,O4}.

\begin{figure*}[htb]
\begin{center}
\includegraphics[width=0.80\textwidth]{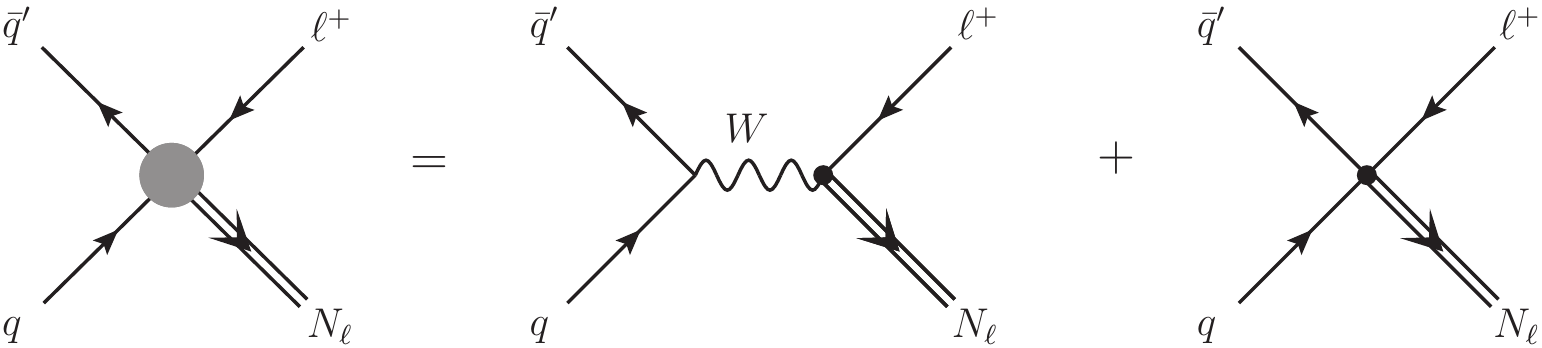}
\end{center}
\caption{Leading order diagrams representing heavy composite Majorana neutrino production. The total interaction is the coherent sum of the gauge and contact interactions. Charge-conjugate reactions are implied. See Ref.~\cite{HCMNpheno}.}
\label{interactions}
\end{figure*}

In this Letter we present, for the first time, the results of a search for heavy composite Majorana neutrinos predicted in the framework of a new model described in Ref.~\cite{HCMNpheno}.
In that reference, the production and decay of $\textrm{N}_{\ell}$ are analyzed, considering both gauge and contact interactions.
The total interaction is given by the coherent sum of the contact and the gauge contributions, as shown in Fig.~\ref{interactions}.
The contribution of the contact interaction to the production cross section is two to three orders of magnitude higher compared to that of the gauge interaction~\cite{HCMNpheno}.

The contact interaction is described by an effective four-fermion Lagrangian of the type
\begin{linenomath}
\begin{equation}
\mathcal{L}_{C} =\frac{g_\ast^2}{\Lambda^2}\frac{1}{2}j^\mu j_\mu ,
\label{eq1}
\end{equation}
\end{linenomath}
with
\begin{linenomath}
\begin{equation}
j_\mu =\eta^a_\textrm{L}\bar{f_\textrm{L}}\gamma_\mu f^\prime_\textrm{L}+\eta^b_\textrm{L}\bar{f^\ast_\textrm{L}}\gamma_\mu f^{\ast\prime}_\textrm{L}+\eta^c_\textrm{L}\bar{f^\ast_\textrm{L}}\gamma_\mu f^\prime_\textrm{L} + \textrm{h.c.} +(\textrm{L}\to \textrm{R}) ,
\label{eq2}
\end{equation}
\end{linenomath}
where $f_\textrm{L}$ and $f_\textrm{L}^\ast$ are the SM and excited left-handed fermion fields, $g_\ast^2=4\pi$, and the $\eta$ factors, which define the chiral structure, are set equal to one.
The gauge interaction between the SM fermions and the excited fermions is described by a magnetic-type coupling
\begin{linenomath}
\begin{equation}
{\cal L}_\textrm{G} =\frac{1}{2 \Lambda} \, \overline{L^\ast_\textrm{R}} \sigma^{\mu\nu} \left(gf \frac{\overrightarrow{\tau}}{2}\cdot \overrightarrow{\textrm{W}}_{\mu\nu} +g'f' Y \textrm{B}_{\mu\nu}\right) L_\textrm{L} + \textrm{h.c.}
\label{eq3}
\end{equation}
\end{linenomath}
where $L^\ast_\textrm{R}$ and $L_\textrm{L}$ are the right-handed excited doublet and left-handed SM doublet,
$g$ and $g^\prime$ are the $\textrm{SU(2)}_\textrm{L}$ and $\textrm{U(1)}_\textrm{Y}$ gauge couplings,
$\overrightarrow{\textrm{W}}_{\mu\nu}$ and $\textrm{B}_{\mu\nu}$ are the field strengths for the $\textrm{SU(2)}_\textrm{L}$ and $\textrm{U(1)}_\textrm{Y}$ gauge fields, respectively, $\overrightarrow{\tau}$ are the Pauli matrices, $Y$ is the weak hypercharge, and
$f$ and $f^\prime$ are dimensionless couplings, assumed to be 1~\cite{Baur}.

\begin{figure}[htb]
\begin{center}
\includegraphics[width=\cmsFigWidth]{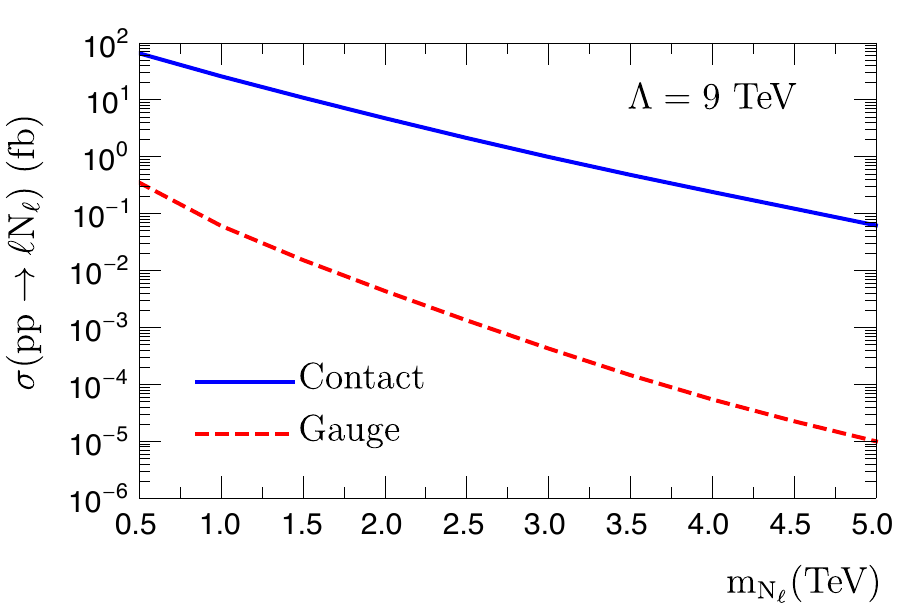}
\includegraphics[width=\cmsFigWidth]{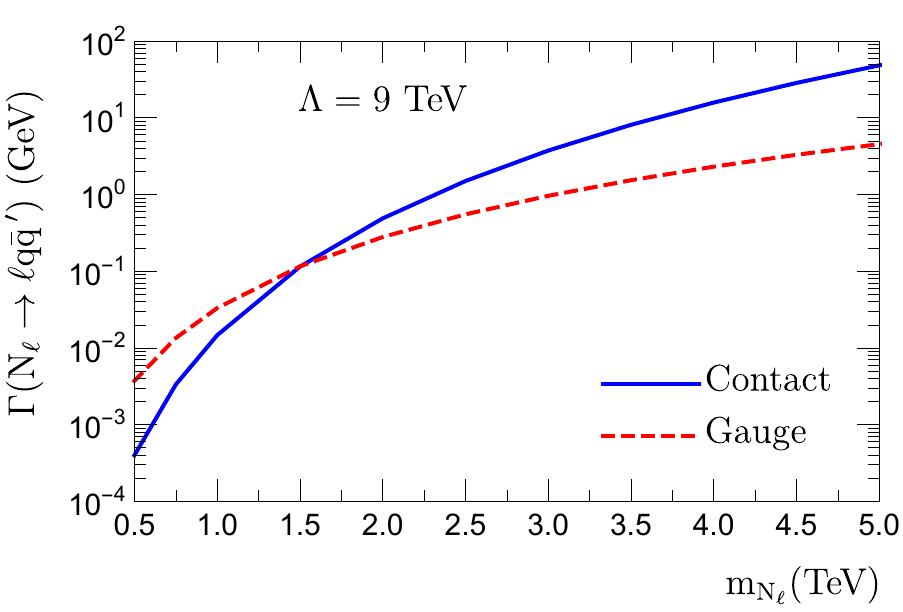}
\end{center}
\caption{Production cross section in pp collisions at $\sqrt{s}= 13\TeV$ of the heavy composite Majorana neutrino via gauge and contact interactions as a function of Majorana mass at $\Lambda = 9\TeV$ (\cmsLeft) and decay width of the heavy composite Majorana neutrino for $\Lambda = 9\TeV $ as a function of its mass (\cmsRight). The figures illustrate LO results of calculations based on Ref.~\cite{HCMNpheno}.
}
\label{xsecdec}
\end{figure}

From the Lagrangians in Eqs.~(\ref{eq1}) and~(\ref{eq3}) we can infer that the higher the value of $\Lambda$, the lower the production cross section of the heavy composite Majorana neutrino.
Since $\textrm{N}_{\ell}$ is its own antiparticle, it can be produced either as a neutrino or an antineutrino. In pp collisions it can be produced in association with a lepton through quark--antiquark annihilation ($\cPq\cPaq^\prime \to \ell^{+} \textrm{N}_{\ell}$). This process can occur via both gauge and contact interactions. The latter is dominant for a wide range of $\Lambda$ values, including the ones to which we are sensitive in this search. Figure~\ref{xsecdec} (\cmsLeft) shows the leading order (LO) production cross section, as a function of the $\textrm{N}_{\ell}$ mass, for the case $\Lambda$ = 9\TeV, which is one of the values considered in this Letter whose calculation is based on Ref.~\cite{HCMNpheno}.

The heavy composite Majorana neutrino can decay through both gauge and contact interactions.
In this case, either the gauge or the contact interaction is dominant, depending on $\Lambda$ and on the mass of $\textrm{N}_{\ell}$, as illustrated in Fig.~\ref{xsecdec} (\cmsRight).

Being a Majorana particle, $\textrm{N}_\ell$ can decay either as a neutrino or an antineutrino with possible decay modes:
\begin{linenomath}
\begin{equation*}
\textrm{N}_{\ell} \to \ell \cPq\cPaq^\prime , \ \ \textrm{N}_{\ell} \to \ell^+\ell^- \nu_{\ell}(\overline{\nu}_{\ell}) , \ \  \textrm{N}_{\ell} \to \nu_{\ell}(\overline{\nu_{\ell}}) \cPq\cPaq^\prime ,
\label{eq4}
\end{equation*}
\end{linenomath}
where the parentheses indicate that the decay product can be a neutrino or an antineutrino.
The possible final states are:
\begin{linenomath}
\begin{equation*}
\ell\ell \cPq\cPaq^\prime , \  \ \ell\ell\ell \nu_{\ell}(\overline{\nu_{\ell}}) ,  \ \  \ell\nu_{\ell}(\overline{\nu_{\ell}}) \cPq\cPaq^\prime .
\label{eq5}
\end{equation*}
\end{linenomath}

In this Letter, the final state
$\ell\ell \cPq\cPaq^\prime$
is considered, as it has the highest sensitivity. We focus on the cases in which $\ell$ is either an electron or a muon, giving rise to the channels
$\Pe\Pe\cPq\cPaq^\prime$ and $\mu\mu \cPq\cPaq^\prime$.

For our analysis we use a data sample of proton-proton collisions at $\sqrt{s}= 13\TeV$ recorded in 2015 with the CMS detector at the CERN LHC, which corresponds to an integrated luminosity of 2.3\fbinv~\cite{REFLUMI}.

Previous searches for compositeness models have been carried out at pp, $\Pp\PAp$, $\Pep\Pem$, and ep colliders.
The most recent results, from the ATLAS and CMS Collaborations, are given in \cite{atlas-limit-new, cms-limit-8TeV}
and exclude the existence of excited electrons (muons) up to masses of 2.45 (2.47)\TeV at 95\% confidence level (CL), for the case $m_{\ell^*} = \Lambda$.
The search performed in the context of Ref.~\cite{HCMNpheno}, which is discussed below, can reach a sensitivity for the existence of heavy composite Majorana neutrinos up to masses of 4.55 (4.77) \TeV for $\textrm{N}_{\Pe}$ ($\textrm{N}_{\mu}$), for the case $m_{\ell^*} = \Lambda$.

Direct searches for heavy neutrinos have been performed by the ATLAS~\cite{ATLAS_HN,ATLAS_HMN} and CMS~\cite{CMS_HN,CMS_HMN,CMS_HMN2,CMS_HNtau} Collaborations. These previous searches have been performed in the $\ell\ell \cPq\cPaq^\prime$ $(\ell = \Pe,\mu,\tau)$ channels,
considering two leptons and two spatially separated jets.
However, in our case, this selection has limited acceptance for gauge boson mediated decays, for which the two jets are expected to overlap,
as they originate from highly Lorentz-boosted hadronic W boson decay products. We overcome this constraint by selecting events with at least
one jet with angular radius large enough to contain a merged pair of partons.
Such a requirement is also highly efficient for heavy composite Majorana neutrino decays mediated by the contact interaction, where we select only one of the two decay jets, as described later in this paper.
This final selection, considered for the first time in a search for heavy neutrinos, could improve the sensitivity of searches for heavy neutrinos in the framework of other models, such as the one considered in Refs.~\cite{CMS_HMN,CMS_HMN2}.

\section{The CMS detector}
\label{The CMS detector}
The central feature of the CMS apparatus is a superconducting solenoid, of 6\unit{m} internal
diameter, providing a field of 3.8\unit{T}. Within the field volume are the inner tracker, the crystal
electromagnetic calorimeter (ECAL), and the brass and scintillator hadron calorimeter (HCAL).
The inner tracker is composed of a pixel detector and a silicon strip tracker, and measures charged-particle trajectories
in the pseudorapidity range $|\eta| < 2.5$.
The finely segmented ECAL consists of nearly $76\,000$ lead-tungstate crystals that provide coverage up to $|\eta| = 3.0$.
The HCAL consists of a sampling calorimeter, which utilizes alternating layers of brass as an absorber and
plastic scintillator as an active material, covering the range $|\eta|< 3$, and is extended to $|\eta|< 5$ by a forward hadron calorimeter.
The muon system covers the region $|\eta| < 2.4$ and consists of up to four planes
of gas ionization muon detectors installed outside the solenoid and sandwiched between the layers of the steel flux-return yoke.
A detailed description of the CMS detector can be found elsewhere~\cite{CMS_DET}.

\section{Data samples and simulation}
\label{The signal and background Monte Carlo samples}

Monte Carlo (MC) event generators are used to simulate the signal and the SM background processes.
The MC samples for the signal are generated at LO with \textsc{CalcHEP} (v3.6)~\cite{CalcHEP_A} for four values of the parameter $\Lambda$: 1, 5, 9, and 13\TeV and
six values of the heavy composite Majorana neutrino mass: 0.5, 1.5, 2.5, 3.5, 4.5, and 6.5\TeV, but only for the cases in which $m_{\textrm{N}_\ell}$ is lower than $\Lambda$. The signal samples produced with $\Lambda = 9\TeV$ are used as reference samples in the analysis, while the samples generated with other values of $\Lambda$ are used to study how the signal efficiency changes, as discussed in Section~\ref{Event selection}.

The simulations for the processes $\ttbar$, $\textrm{tW}$, and $\PAQt\textrm{W}$ (the latter two referred as $\textrm{tW}$ in the rest of the paper) are performed at next-to-the-leading order (NLO) with \POWHEG{} (v2)~\cite{Nason:2004rx,powheg1,powheg2}, while the Drell--Yan (DY) and the W+jets samples are generated with \MADGRAPH{}5\_a\MCATNLO (v5.2)~\cite{Alwall:2014hca}. The WW, WZ, and ZZ processes are produced with \PYTHIA{} (v8.2)~\cite{PYTHIA_82} and are normalized to NLO.

The {NNPDF} 3.0~\cite{NNPDF3} parton distribution functions (PDF) are used, and all simulated samples use the \PYTHIA{} program with the  {CUETP8M1} tune~\cite{CUETP8M1} to describe parton showering
and hadronization.
Additional collisions in the same or adjacent bunch crossings (pileup) are taken into account
by superimposing simulated minimum bias interactions onto the hard scattering process,
with a number distribution matching that observed in data. Simulated events are propagated through
the full \GEANTfour based simulation~\cite{GEANT} of the CMS detector.

\section{Event selection}
\label{Event selection}
Single-lepton triggers that require either an electron with transverse momentum $\pt > 105\GeV$ or a muon with $\pt > 50\GeV$ within $|\eta| < 2.4$, are thus used to select events in the $\Pe\Pe \cPq\cPaq^\prime$ and $\mu\mu \cPq\cPaq^\prime$ channels, respectively.
As the signal is characterized by high-momentum leptons in the final state, the difference in trigger thresholds does not affect the relative signal sensitivity.

Electrons are reconstructed as superclusters in the ECAL associated with tracks in the tracking detector~\cite{electron8TeV}.
Requirements on energy deposits in the calorimeter and number of track measurements are imposed to distinguish prompt electrons from charged pions and from electrons produced by photon conversions.
Muons are reconstructed using the inner tracker and muon detectors~\cite{muon7TeV}.
Quality requirements, based on the minimum number of measurements in the silicon tracker, pixel detector, and muon detectors
are applied in order to suppress backgrounds from decays in flight of hadrons and
from hadron shower remnants that reach the muon system.
We require exactly two electrons or exactly two muons and the two leptons must come from the same vertex.
The \pt of the leading (subleading) electron is required to be higher than 110 (35)\GeV;
the corresponding threshold for muons is 53 (30)\GeV. All lepton candidates have to be reconstructed within $|\eta| < 2.4 $ and to pass isolation requirements, as specified in~\cite{electron8TeV,muon7TeV} in order to reduce background from misidentified jets.

Jets are reconstructed using the anti-\kt clustering algorithm~\cite{Jet1,Jet2}
applied to the objects reconstructed with the particle-flow (PF) algorithm~\cite{CMS_PF}.
The latter combines information from all CMS subdetectors and reconstructs individual particles in the event (electrons, muons, photons, neutral and charged hadrons).
Jets are reconstructed with a distance parameter of $R = 0.8$, and are referred to as ``large-radius jets" (and labelled by the symbol ``J") in the rest of the text. This distance parameter is suitable for reconstructing jets that originate from both gauge and contact interactions.
The large-radius jets are required to have a $ \pt > 190\GeV$, $|\eta|< 2.4$, and to be separated from leptons
by $\Delta R = \sqrt{\smash[b]{(\Delta \eta)^{2} + (\Delta \phi)^{2}}} > 0.8$, where $\phi$ is the azimuthal angle.

Using MC signal samples we find that requiring one or more large-radius jets guarantees high signal efficiency for events with two leptons (greater than 95\% for heavy composite Majorana neutrinos of masses above 1\TeV) and is suitable for $\textrm{N}_{\ell}$ decays through both the gauge and the contact interactions.
The signal region for the search for heavy composite Majorana neutrinos is defined by requiring two leptons, selected without specifying the charge, with invariant mass $m_{\ell \ell}> 300\GeV$ and at least one large-radius jet satisfying the previously described requirements. The requirement on $m_{\ell \ell}$ is introduced in order to reduce the DY background and part of the $\ttbar$ background, without affecting the signal acceptance. With this selection, the total efficiency for events entering the signal region is expected to be about 50\% in the $\Pe\Pe\cPq\cPaq^\prime$ channel and 75\% in the $\mu\mu \cPq\cPaq^\prime$ channel, for masses of $\textrm{N}_{\ell}$ greater than 1\TeV and $\Lambda = 9\TeV$.
We find that the total signal efficiency varies by at most 25\% for signal samples produced with $\Lambda$ of 1 and 13\TeV, while it changes less than 3\% for signal samples generated with $\Lambda$ between 5 and 13\TeV.
A shape-based analysis is performed, searching for evidence of a signal by considering the distribution of the mass of the two leptons and the leading large-radius jet, $m_{\ell \ell \mathrm{J}}$. This variable provides good discrimination between the signal and the SM background contributions, as can be seen in Fig.~\ref{fig:UnblidedMasses}.

\section{Background estimation}
\label{Background estimation}
\begin{figure*}
\begin{center}
\includegraphics[width=\cmsFigWidth]{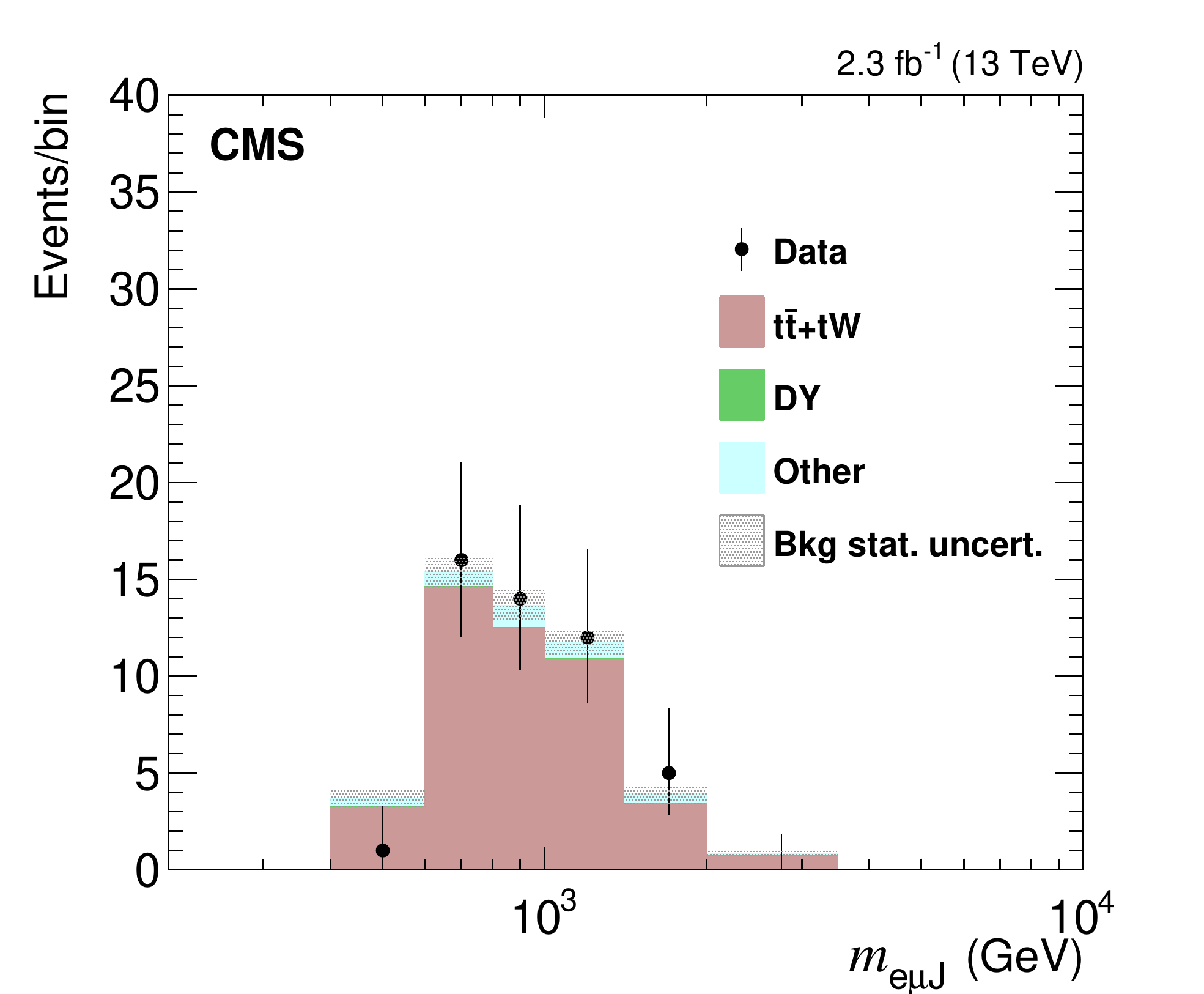}
\includegraphics[width=\cmsFigWidth]{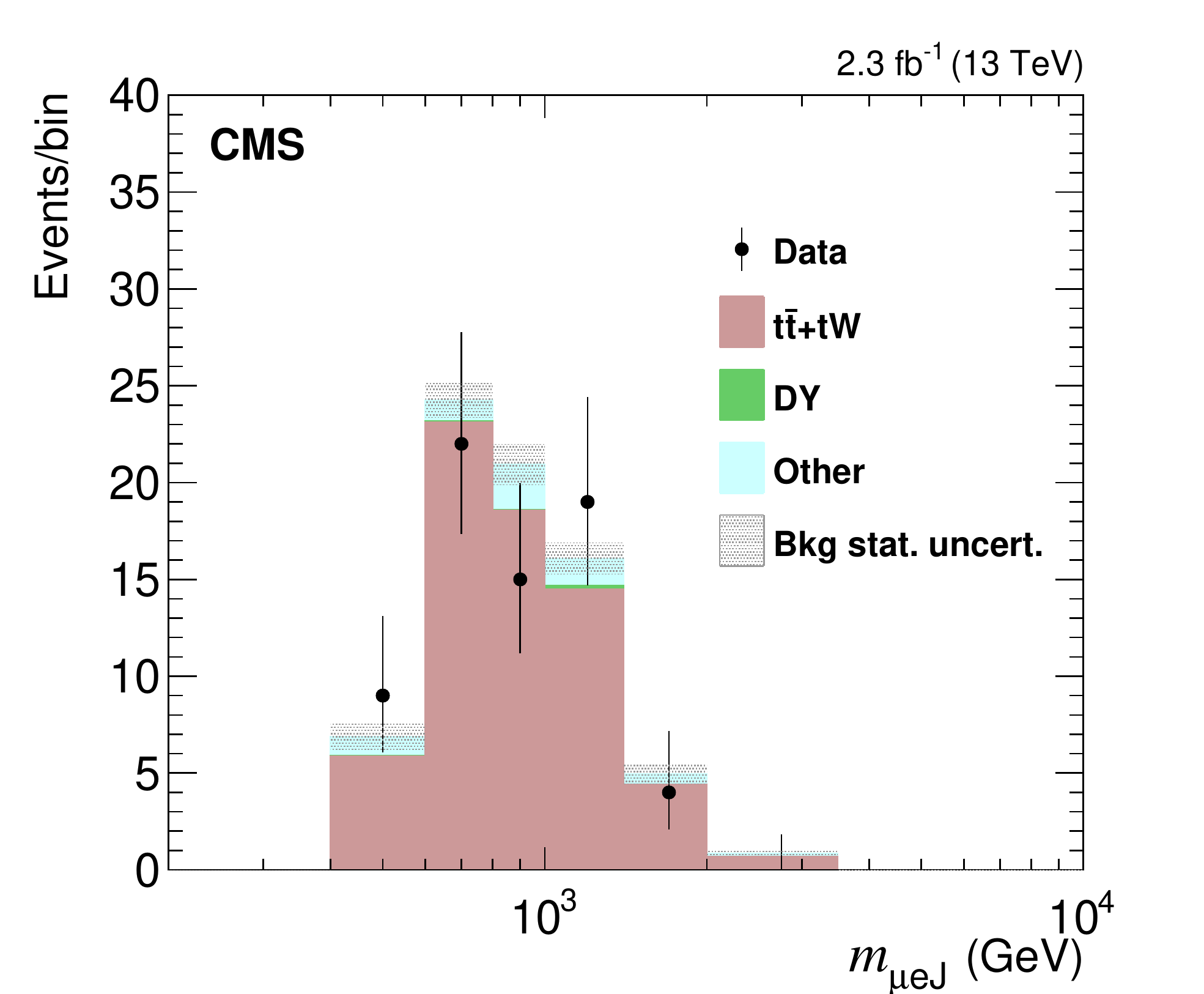}
\caption{Data events in the $\Pe\mu \cPq\cPaq^\prime$ control regions used to estimate the $\ttbar + \textrm{tW}$ contribution in the $\Pe\Pe\cPq\cPaq^\prime$ (left) and $\mu\mu \cPq\cPaq^\prime$ (right) channels, compared to the expectations of the background simulations.
``Other" stands for the contribution from W+jets and diboson events.
}
\label{fig:EMuDataMC}
\end{center}
\end{figure*}

The dominant background process is top quark pair production, $\ttbar$, which is estimated together with the single top quark $\textrm{tW}$ contribution. These two sources of background are always considered together in this analysis and the combination is referred to as $\ttbar + \textrm{tW}$.
The estimation is performed using a control sample in data free of signal contamination. This control sample consists of
$\ell\ell^\prime \cPq\cPaq^\prime$ events in which
one lepton is required to be an electron and the other a muon. The acceptance and
efficiency differ between muons and electrons, and we find that
the overall event efficiency depends most strongly on the kinematics of
the leading lepton. We therefore define samples of separate and distinct final states $\Pe\mu \mathrm{J}$ and $\mu \Pe\mathrm{J}$,
where the first named lepton is the leading one, as control samples
for the signal samples $\Pe\Pe\mathrm{J}$ and $\mu\mu \mathrm{J}$, respectively. All other
requirements are the same for both the control and signal samples.
We have verified that the $m_{\Pe\Pe\mathrm{J}}$ and $m_{\mu \mu \mathrm{J}}$ distributions are well-modelled
by the corresponding $m_{\Pe\mu \mathrm{J}}$ and $m_{\mu \Pe\mathrm{J}}$ distributions using the MC samples.
Figure~\ref{fig:EMuDataMC} shows good agreement between data and expectation from MC simulation for the $m_{\Pe\mu \mathrm{J}}$ and the $m_{\mu \Pe\mathrm{J}}$ distributions.
Backgrounds from processes other than $\ttbar + \textrm{tW}$ in these control samples are estimated from simulation and subtracted prior to being transferred to the signal region.
The final $\ttbar + \textrm{tW}$ contribution is estimated from the mass
shapes of the different flavour control regions scaled to the signal regions by transfer factors.
The transfer factors depend on $m_{\ell\ell J}$ and are estimated in bins corresponding to those of Fig.~\ref{fig:EMuDataMC}, which are then used for the final mass distributions in the signal regions. They are evaluated
using MC simulation and account for differences in acceptances and efficiencies of selected
$\Pe\mu \cPq\cPaq^\prime$ and $\mu \Pe\cPq\cPaq^\prime$ events of the control regions with respect to
the selected $\Pe\Pe\cPq\cPaq^\prime$ and $\mu\mu \cPq\cPaq^\prime$ events of the signal regions.

The DY process gives rise to another source of background when two leptons are produced together with initial-state radiation that results in a jet. This contribution is estimated from the MC simulation normalised to data in the signal-free region around the Z boson mass peak given by $80 <M(\ell \ell)< 100\GeV$.
In order to check the validity of the measured normalization factors for masses above the Z boson mass peak, we compare the data with the MC prediction in the signal-depleted region given by $ 100 <M(\ell \ell)< 300\GeV$. We find that the normalisation factors vary by 8\% between these two mass ranges, and we use this value to assign a systematic uncertainty. The statistical and systematic uncertainties of the normalisation factor are then combined with the statistical uncertainty of the DY simulation to estimate the total systematic uncertainty.

Multijet events with at least three jets may enter the signal or control region for estimating backgrounds related to top quarks
if two of these jets are misidentified as leptons.
The contamination due to this process is found to be negligible because of the low rate at which jets are misidentified as leptons. We verify this with a method developed in the CMS search for $Z'$-like resonances in electron pair or muon pair final states~\cite{ZpCMSmumuee13}.
Nonisolated lepton candidates selected from data are weighted by a correction
factor to extrapolate the final contribution to the signal region. The correction factor is the rate of a nonisolated lepton to pass the full selection and is measured from data as a function of $\pt$ and~$\eta$.

The other SM backgrounds, arising from W+jets and diboson production, are small (${\sim}5\%$ of the total), and their contribution is taken directly from MC simulation.

\section{Systematic uncertainties}
\label{Systematic uncertainties}
The systematic uncertainties are taken into account through their effect on the mass distribution and the yield normalisation.
The uncertainty in the calculation of the  $\ttbar + \textrm{tW}$ and DY backgrounds is dominated by the statistical uncertainty of the control samples used for the estimations. The contamination from sub dominant backgrounds in these control regions has a negligible effect on the systematic uncertainty. The uncertainties related to the background estimations vary between 20 and 30\% (40 and 100\%) from the lowest to the highest mass bin in which the $\ttbar + \textrm{tW}$ (DY) processes contribute. The uncertainty associated with the estimation of the W+jets and diboson backgrounds, which together represent only a small fraction of the events in the signal region, has a negligible impact on the limit calculation.
The uncertainty in the acceptance associated with the PDFs is evaluated in accordance with the PDF4LHC recommendations~\cite{PDF4LHC},
using the PDF4LHC15 Hessian PDF set with 100 eigenvectors. The PDF uncertainty amounts to 10\% for the DY background and 4\% for the signal.
Uncertainties related to the trigger, and to lepton reconstruction, identification, and isolation efficiencies
are measured by dedicated analyses using $ \PZ \to \ell \ell$ events~\cite{electron8TeV,muon7TeV} and
amount on average to 3\% (6\%) for the background and 4\% (10\%) for the signal in the electron (muon) channel.
The systematic uncertainty in the lepton energy scale and resolution are found to be approximately 5\% (6\%) for the background and 3\% (4\%) for the signal.
The uncertainty related to jet energy scale amounts to 3\% (4\%) for the background of the $\Pe\Pe\cPq\cPaq^\prime$ ($\mu\mu \cPq\cPaq^\prime$) channel,
and around 1\% for the signal regardless of the channel.
Uncertainties related to jet energy resolution correspond to 2 and 4\% for the background and the signal, respectively, in both channels.
The imperfect modelling of pileup interactions leads to a systematic uncertainty of about 4\% for background and 2\% for signal.
The uncertainty in the total integrated luminosity amounts to 2.3\% for the 2015 data~\cite{REFLUMI}.

\section{Results}
\label{Results}
Table~\ref{tab:SRevt} lists the estimated background yields, the total number of observed events for each channel and the number of events expected for the signal, considering $\Lambda = 9\TeV$ and two hypotheses for the masses of $\textrm{N}_{\ell}$: 1.5 and 2.5\TeV.
Table~\ref{tab:SRevt} (upper) shows the number of events integrated over all values of the reconstructed mass,
while Table~\ref{tab:SRevt} (lower) shows the agreement for the high sensitivity region above 1.4 \TeV.

\begin{figure*}[htbp]
\centering
\includegraphics[width=\cmsFigWidth]{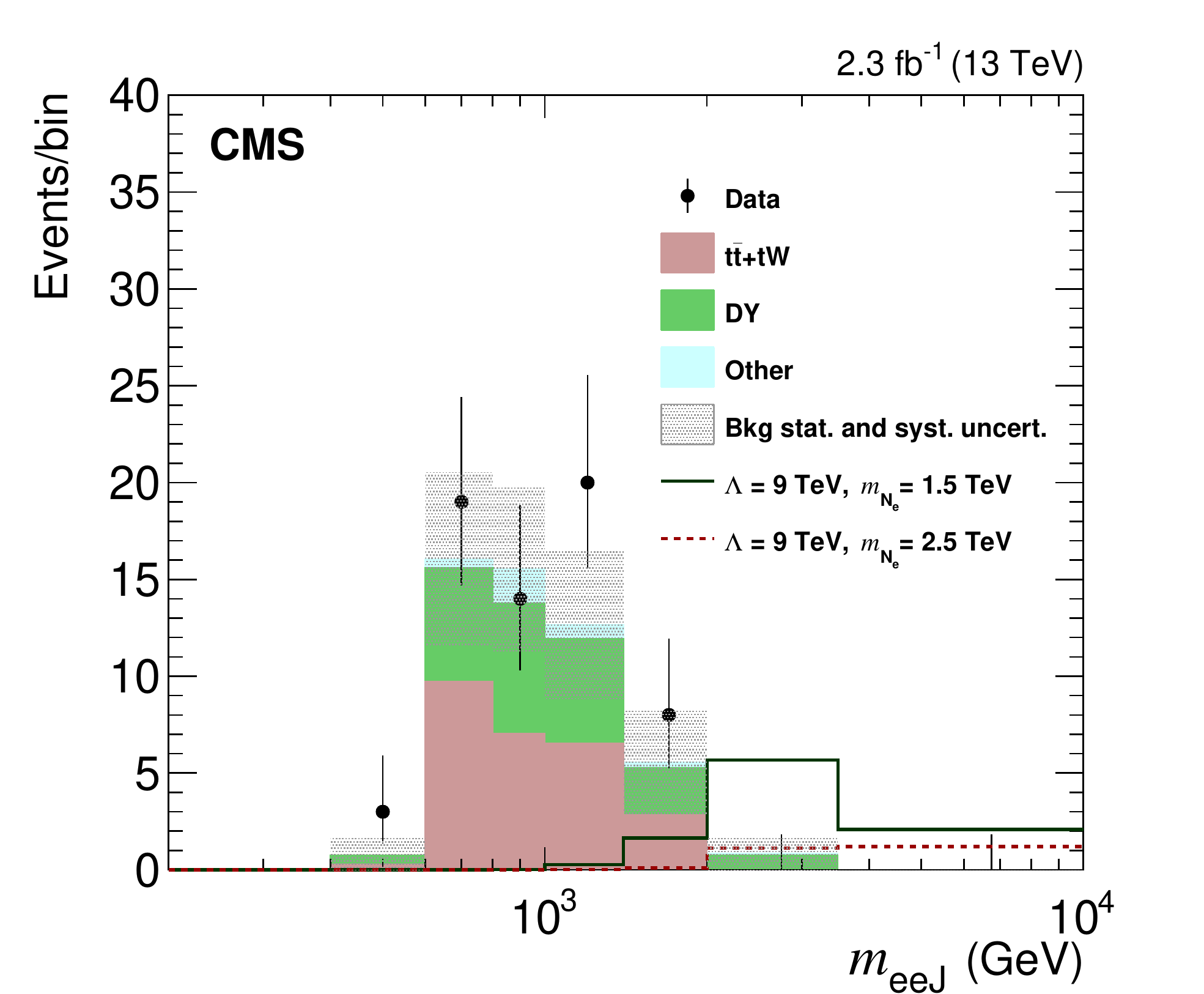}
\includegraphics[width=\cmsFigWidth]{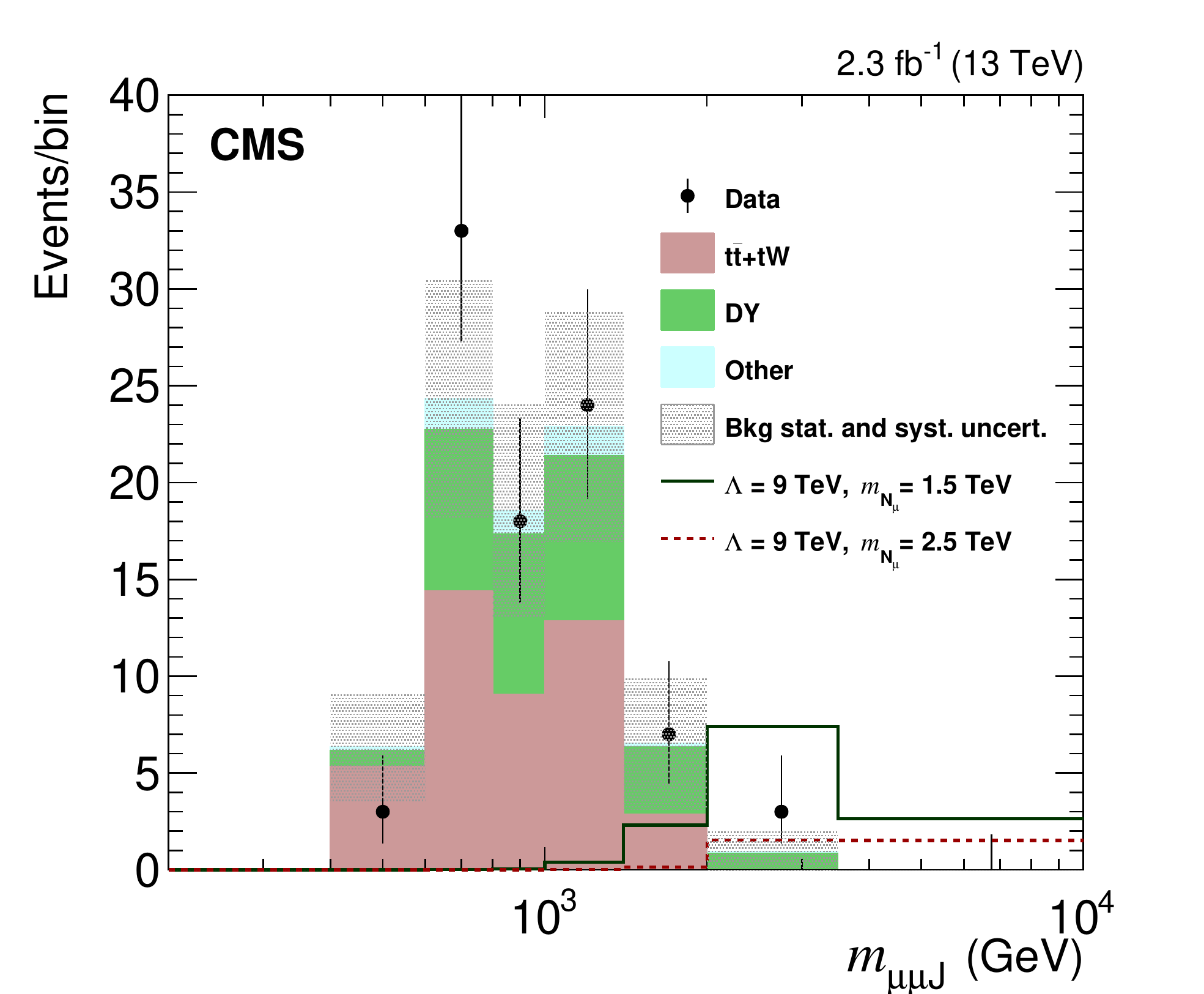}
\caption{Distribution of the variable $m_{\ell \ell \mathrm{J}}$ for the data (black points), the estimated SM backgrounds (stacked filled histograms), and the signal (lines) with $\Lambda$ = 9\TeV and masses of $\textrm{N}_{\ell}$ equal to 1.5 and 2.5\TeV, for the $\Pe\Pe\cPq\cPaq^\prime$ (left) and the $\mu \mu \cPq\cPaq^\prime$ (right) channels. ``Other" stands for the contribution from W+jets and diboson events. The background uncertainties are the combined statistical and systematic uncertainties.}
\label{fig:UnblidedMasses}
\end{figure*}

Distributions of $m_{\ell \ell \mathrm{J}}$ are shown in Fig.~\ref{fig:UnblidedMasses}, where the data are compared with the estimated SM backgrounds and two different signal hypotheses are superimposed. The figures are for the $\Pe\Pe\cPq\cPaq^\prime$ (left) and $\mu \mu \cPq\cPaq^\prime$ (right) channels. The uncertainties on the background estimation are the combination of the statistical and the systematic uncertainties.

\begin{table*}[h]
\topcaption{Number of events observed in data are compared to the expected background yields and those of a hypothetical heavy composite Majorana neutrino of mass 1.5 and 2.5\TeV, and $\Lambda$ = 9\TeV, given for all values of $m_{\ell \ell \mathrm{J}}$ (upper table) and for $m_{\ell \ell \mathrm{J}}>1.4$\TeV (lower table). The expected signal yields are computed at LO accuracy. ``Other" stands for the contribution from W+jets and diboson events. The background and signal simulation yields are given with both statistical and systematic uncertainties. Statistical uncertainties given as 0.0 correspond to values much smaller than the systematical uncertainty.}
\begin{center}
\begin{tabular}{ccc}
\hline
\multirow{2}{*}{Process (all $m_{\ell \ell \mathrm{J}}$)} & $\Pe\Pe\cPq\cPaq^\prime$   &$\mu\mu \cPq\cPaq^\prime$ \\
 & $(\text{mean} \pm \text{stat}  \pm \text{syst})\x$ & $(\text{mean} \pm \text{stat}  \pm \text{syst})\x$  \\
\hline
$\ttbar+\textrm{tW}$          &   $ 26  \, \pm \,  \ 4 \   \, \pm \,  3\x $ &  $ 44  \, \pm \,  \ 6 \  \, \pm \,  5\x $ \\
Drell\--Yan              &  $ 22  \, \pm \,  \ 1 \  \, \pm \,  5\x $ &  $ 30  \, \pm \, \ 1  \  \, \pm \,  7\x $ \\
Other                     &  $ 3.3  \, \pm   0.8   \pm \,  0.1 $ &  $ 4.7  \, \pm   0.9  \pm \,  0.4 $ \\
\hline
Total                   &  $ 51  \, \pm \,  \ 4 \  \, \pm \,  6\x $ &  $ 80  \, \pm \,  \ 6  \  \, \pm \,  8\x $ \\
\hline
Observed               &  $ 64$ & $  88  $ \\
\hline
$\textrm{N}_{\ell}$ (1.5\TeV )  & $ 9.7  \, \pm   0.0   \pm \,  0.3$ & $ 12.8  \, \pm   0.0   \pm \,  1.6\x $ \\
$\textrm{N}_{\ell}$ (2.5\TeV )  & $ 2.4  \, \pm  0.0   \pm \,  0.1$ & $ 3.2  \, \pm   0.0   \pm \,  0.4 $ \\
\hline
&  & \\
\end{tabular}
\begin{tabular}{ccc}
\hline
\multirow{2}{*}{Process ($m_{\ell \ell \mathrm{J}} >$ 1.4\TeV) } & $\Pe\Pe\cPq\cPaq^\prime$   &$\mu\mu \cPq\cPaq^\prime$ \\
 & $(\text{mean} \pm \text{stat}  \pm \text{syst})\x$ & $(\text{mean} \pm \text{stat}  \pm \text{syst})\x$  \\
\hline
$\ttbar+\textrm{tW}$              & $ 2.8 \, \pm  1.5  \pm \, 0.9$ & $2.9 \, \pm  1.8  \pm \, 1.3$ \\
Drell--Yan                  & $ 3.2 \, \pm  0.3  \pm \, 2.0$ & $4.3 \, \pm  0.4  \pm \,2.7$ \\
Other                      & $ 0.36 \, \pm  0.10  \pm \, 0.04$ & $0.25 \, \pm  0.10  \pm \, 0.11 $ \\
\hline
Total                      & $ 6.4 \, \pm  1.5  \pm \, 2.2$ & $7.5 \, \pm  1.8  \pm \, 3.0 $ \\
\hline
Observed                   & $ 8 $ & $ 10 $  \\
\hline
$\textrm{N}_{\ell}$ (1.5\TeV ) & $ 9.4 \, \pm  0.0  \pm \, 0.3$ & $12.4 \, \pm  0.0  \pm \, 1.6\x $ \\
$\textrm{N}_{\ell}$ (2.5\TeV ) & $ 2.4 \, \pm  0.0  \pm \, 0.1$ & $3.2 \, \pm  0.0  \pm \, 0.4 $ \\
\hline
\end{tabular}
\end{center}
\label{tab:SRevt}
\end{table*}

The observations are in agreement with the background expectations from the SM.
We use a modified frequentist $CL_s$ criterion~\cite{CLs,Junk} to set an upper limit at 95\% CL on the product cross section for production of the heavy composite Majorana neutrino produced in association with a lepton and the branching fraction for the $\textrm{N}_{\ell}$ decay to a same-flavour lepton and two quarks, $\,(\Pp\Pp \to \ell \textrm{N}_{\ell}) \, {\cal B}(\textrm{N}_{\ell} \to \ell \cPq\cPaq^\prime)$. We also set upper limits on the compositeness scale $\Lambda$. The $m_{\ell \ell \mathrm{J}}$ distributions for the MC signal, SM backgrounds, and observed data are used as input in the limit computation together with the systematic uncertainties discussed in Section~\ref{Systematic uncertainties}, which are treated as uncorrelated among the bins of the mass distribution, if they are related to the background estimations, and correlated otherwise.

\begin{figure*}[htb]
\begin{center}
\includegraphics[width=\cmsFigWidth]{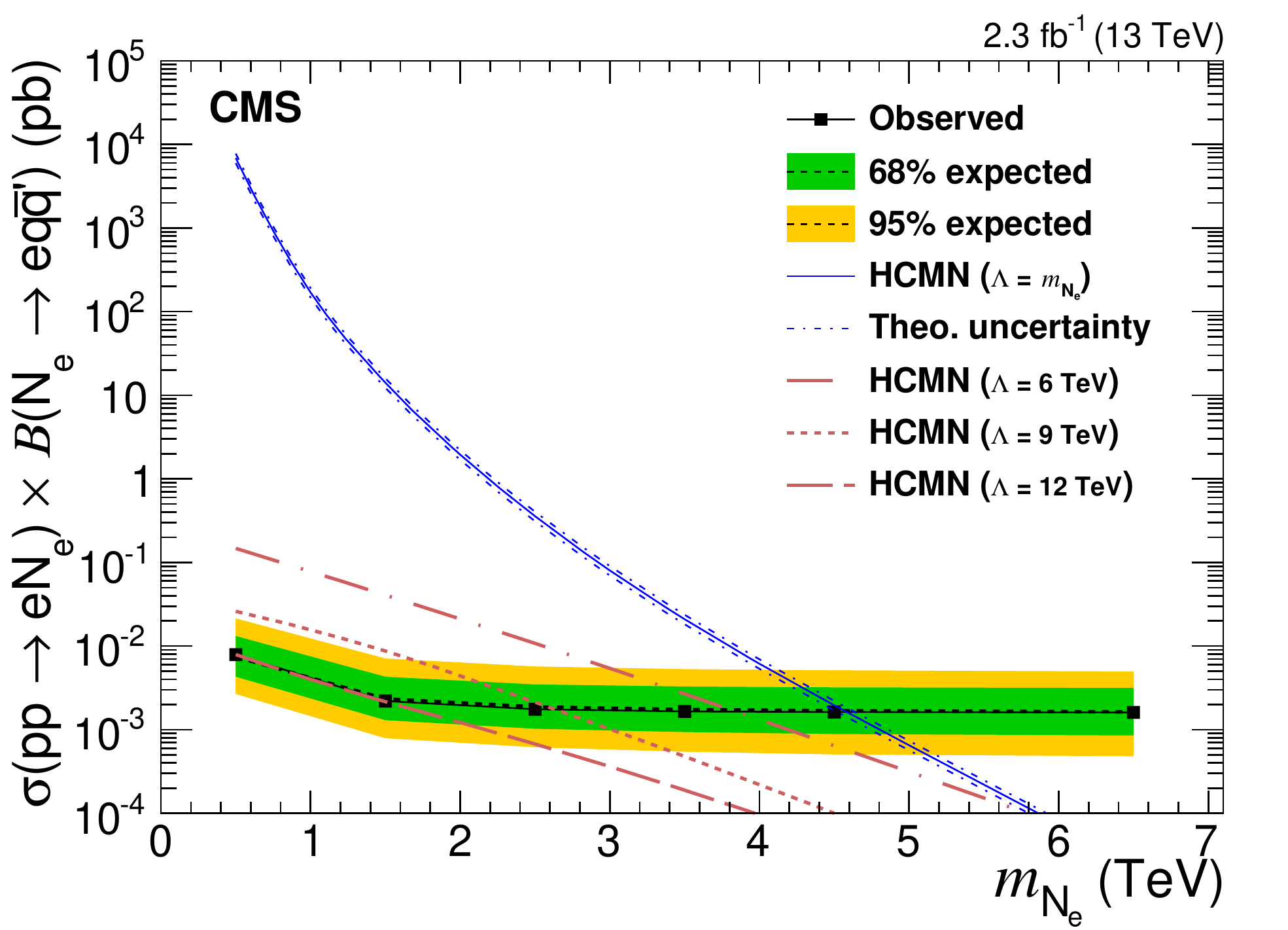}
\includegraphics[width=\cmsFigWidth]{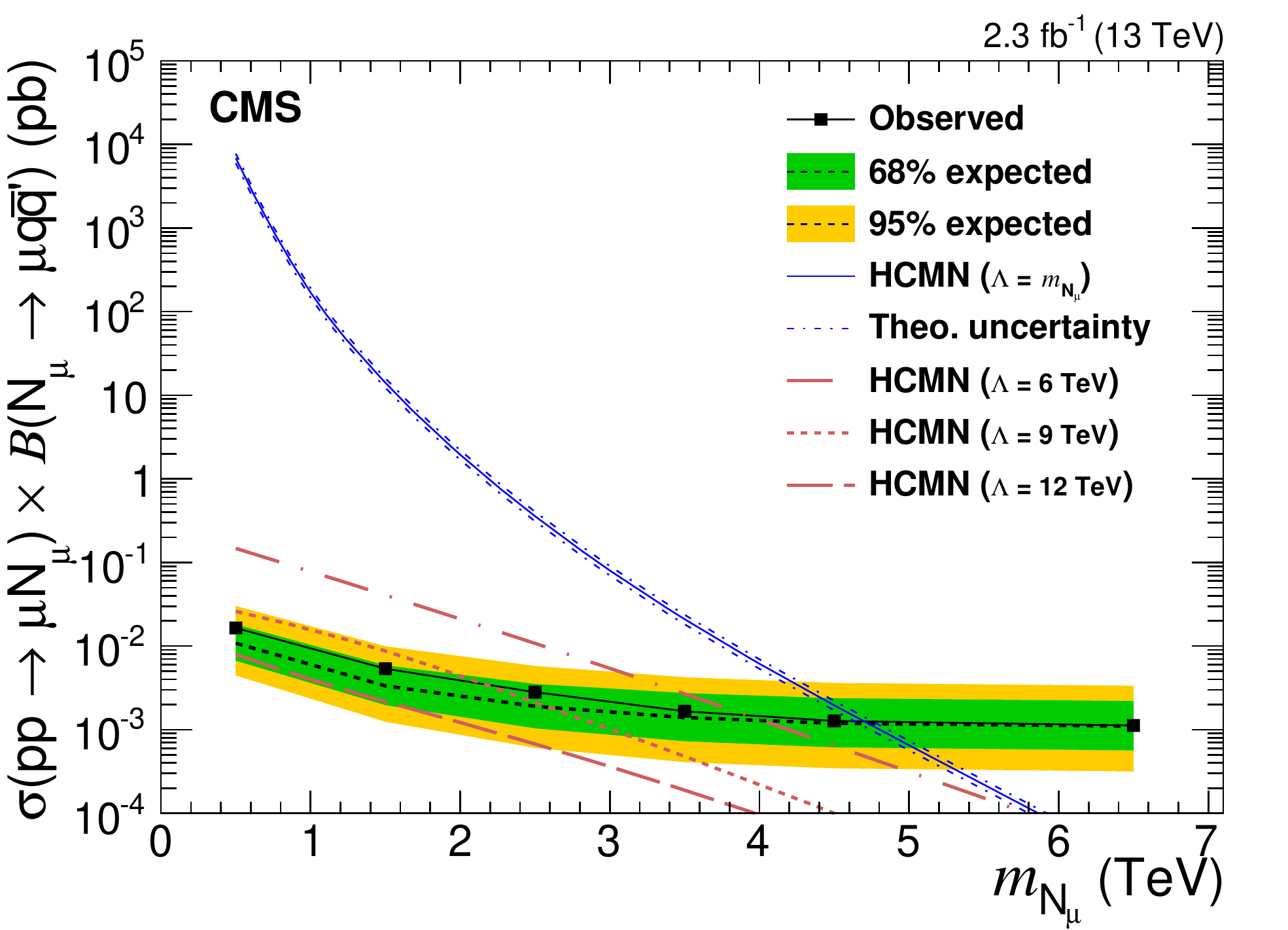}
\end{center}
\caption{The observed 95\% CL upper limits (solid black lines) on $\sigma(\Pp\Pp \to \ell \textrm{N}_{\ell}) \, {\cal B}(\textrm{N}_{\ell} \to \ell \cPq\cPaq^\prime)$, obtained in the analysis of the $\Pe\Pe\cPq\cPaq^\prime$ (left) and the $\mu\mu \cPq\cPaq^\prime$ (right) final states, as a function of the mass of the heavy composite Majorana neutrino, $\textrm{N}_{\ell}$. The corresponding expected limits are shown by the dotted lines, and the bands represent the expected variation of the limit to one and two standard deviation(s).
The solid blue curve indicates the theoretical prediction of $\sigma(\Pp\Pp \to \ell \textrm{N}_{\ell}) \, {\cal B}(\textrm{N}_{\ell} \to \ell \cPq\cPaq^\prime)$ for $\Lambda = m_{\textrm{N}_{\ell}}$. The uncertainty in the theoretical prediction is derived by taking into account the factorizatin and normalization scales. The light red textured curves give the theoretical predictions for three $\Lambda$ values ranging from 6 to 12\TeV in steps of 3\TeV.
}
\label{Limit_eeqq_mumuqq}
\end{figure*}

\begin{figure*}[htb]
\begin{center}
\includegraphics[width=\cmsFigWidth]{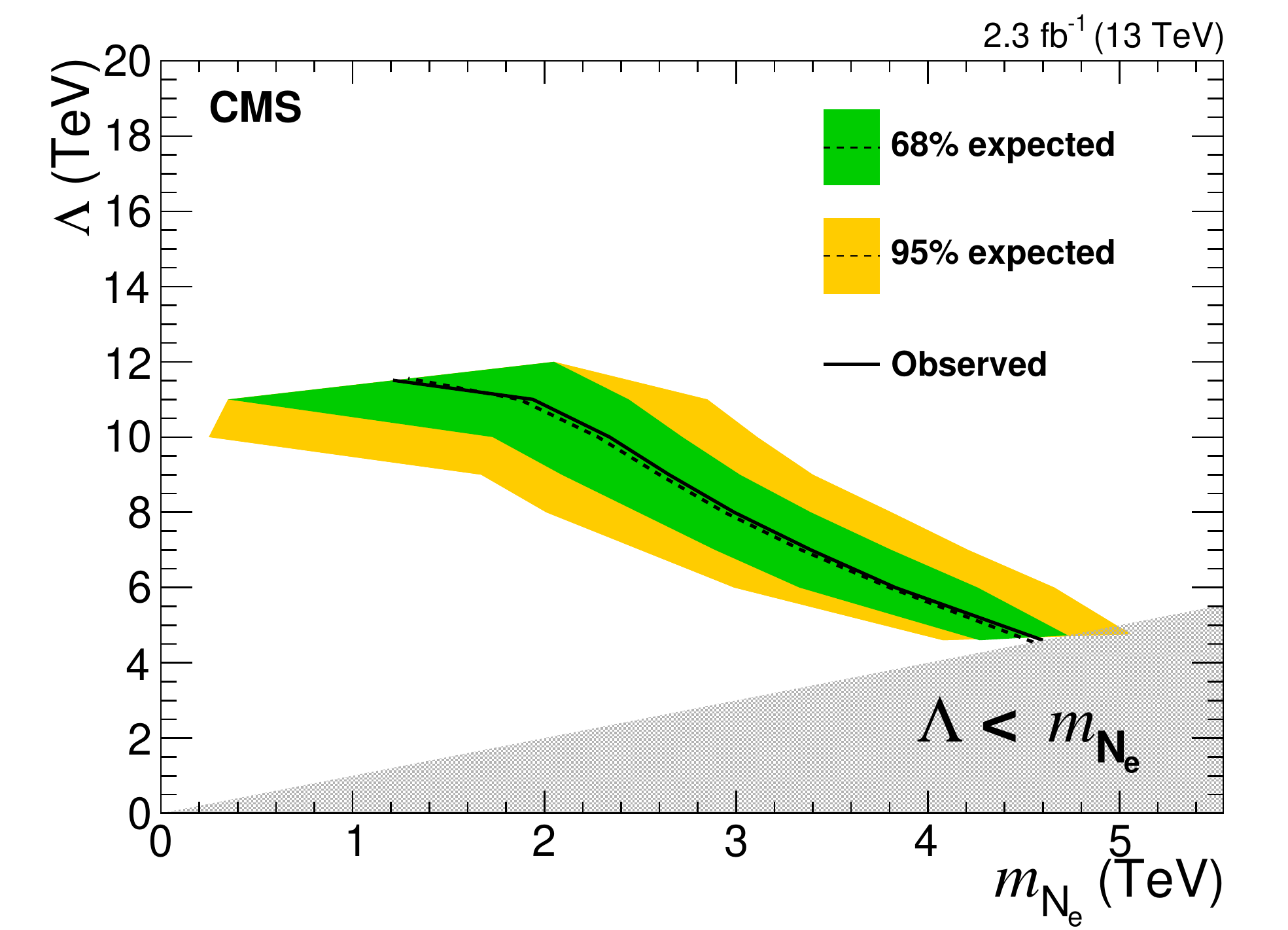}
\includegraphics[width=\cmsFigWidth]{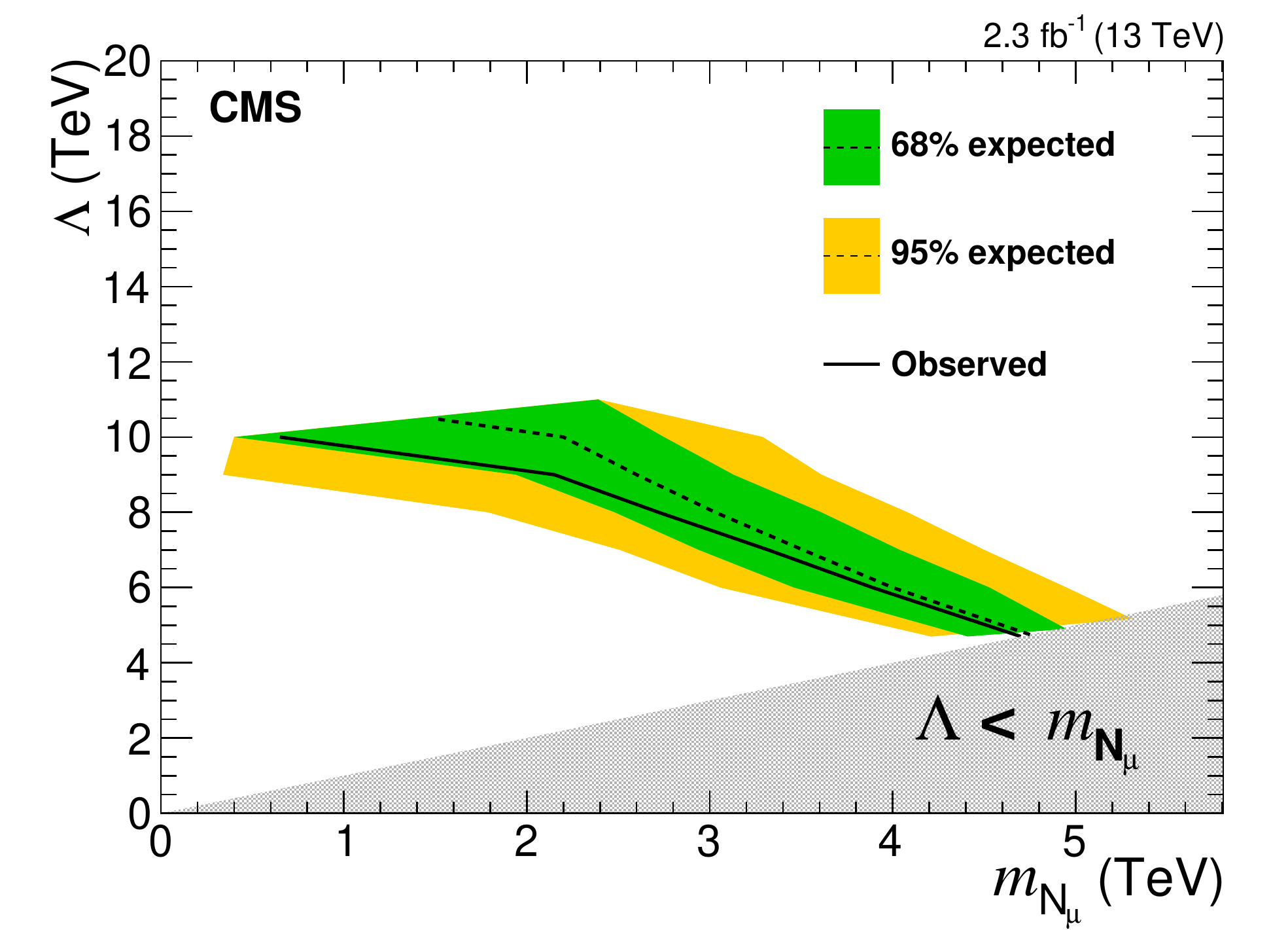}
\end{center}
\caption{
The observed 95\% CL lower limits (solid black lines) on the compositeness scale $\Lambda$, obtained in the analysis of the $\Pe\Pe\cPq\cPaq^\prime$ (left) and the $\mu\mu \cPq\cPaq^\prime$ (right) final states, as a function of the mass of the heavy composite Majorana neutrino, $\textrm{N}_{\ell}$. The dotted lines represent the corresponding expected limits and the bands, the expected variation to one and two standard deviation(s).
The grey zone represents the phase space $\Lambda < M_{\textrm{N}_{\ell}}$, which is not allowed by the model.
}
\label{Limit_lambda_mass}
\end{figure*}

The observed and expected upper limits on $\sigma(\Pp\Pp \to \ell \textrm{N}_{\ell}) \, {\cal B}(\textrm{N}_{\ell} \to \ell \cPq\cPaq^\prime)$ as a function of the mass of the heavy composite Majorana neutrino
are shown in Fig.~\ref{Limit_eeqq_mumuqq}.
The bands represent expected variations of the limit to one and two standard deviation(s).
The solid blue curve indicates the theoretical prediction of $\sigma(\Pp\Pp \to \ell \textrm{N}_{\ell}) \, {\cal B}(\textrm{N}_{\ell} \to \ell \cPq\cPaq^\prime)$ for $m_{\textrm{N}_{\ell}} = \Lambda$, while the light red textured curves show the same theoretical prediction for three $\Lambda$ values ranging from 6 to 12\TeV.
The corresponding exclusion limits on the compositeness scale $\Lambda$ are displayed in Fig.~\ref{Limit_lambda_mass}.
At low $\textrm{N}_{\ell}$ masses, values of the compositeness scale $\Lambda$ can be excluded up to 11.5 and 10.0\TeV in the $\Pe\Pe\cPq\cPaq^\prime$ and $\mu\mu \cPq\cPaq^\prime$ channel, respectively.
The sensitivity to $\Lambda$ decreases at higher masses of $\textrm{N}_{\ell}$. For the case of $m_{\textrm{N}_{\ell}} = \Lambda$, the resulting exclusion limits on $m_{\textrm{N}_{\ell}}$ are up to 4.60 (4.55)\TeV in the $\Pe\Pe\cPq\cPaq^\prime$ channel and 4.70 (4.75)\TeV in the $\mu\mu \cPq\cPaq^\prime$ channel, considering the observation (SM expectation).
When deriving these limits, we assume that the signal efficiency is independent of $\Lambda$. This hypothesis has been validated for signal samples produced with $\Lambda$ between 5 and 13\TeV, while for samples with $\Lambda$ lower than 5\TeV the difference can be up to 25\%. Despite this difference, the whole region in Fig.~\ref{Limit_lambda_mass} remains excluded because of the much higher cross section for lower $\Lambda$ points.
We further verify that the upper limits on $m_{\textrm{N}_{\ell}}$ for a given $\Lambda$ value vary at most by 5\% comparing the cases in which the MC signal $m_{\ell \ell \mathrm{J}}$ distributions produced with $\Lambda$ equal to 5 and 13 TeV are used as input in the limit calculation.

\section{Summary}
\label{Summary}
A search for physics beyond the standard model has been performed in the framework of a new model~\cite{HCMNpheno} predicting a heavy Majorana neutrino, $\textrm{N}_{\ell}$, that originates from a composite-fermion scenario and is produced in association with a matched-flavour charged lepton.
The measurement is performed analysing the final state with two leptons, selected without specifying the charge, and at least one large-radius jet, a signature not previously utilized in searches for heavy neutrinos.
The data set used corresponds to an integrated luminosity of 2.3 fb$^{-1}$ collected with the CMS detector at the LHC in pp collisions at $\sqrt{s}$ = 13\TeV. Good agreement between the data and the standard model prediction is observed.
Upper limits are set at 95\% confidence level both on the product of the cross section $\sigma(\Pp\Pp \to \ell \textrm{N}_{\ell})$
and the branching fraction ${\cal B}(\textrm{N}_{\ell} \to \ell \cPq\cPaq^\prime)$, and on the compositeness scale $\Lambda$, as a function of $m_{\textrm{N}_{\ell}}$, $\ell$ being an electron or a muon.
For the representative case $\Lambda = m_{\textrm{N}_{\ell}}$, $\textrm{N}_{\Pe}$ masses up to 4.60 \TeV and $\textrm{N}_{\mu}$ masses up to 4.70 \TeV are excluded.
This measurement represents the first search that places constraints on the model described in Ref.~\cite{HCMNpheno}.

\begin{acknowledgments}

We congratulate our colleagues in the CERN accelerator departments for the excellent performance of the LHC and thank the technical and administrative staffs at CERN and at other CMS institutes for their contributions to the success of the CMS effort. In addition, we gratefully acknowledge the computing centres and personnel of the Worldwide LHC Computing Grid for delivering so effectively the computing infrastructure essential to our analyses. Finally, we acknowledge the enduring support for the construction and operation of the LHC and the CMS detector provided by the following funding agencies: BMWFW and FWF (Austria); FNRS and FWO (Belgium); CNPq, CAPES, FAPERJ, and FAPESP (Brazil); MES (Bulgaria); CERN; CAS, MoST, and NSFC (China); COLCIENCIAS (Colombia); MSES and CSF (Croatia); RPF (Cyprus); SENESCYT (Ecuador); MoER, ERC IUT, and ERDF (Estonia); Academy of Finland, MEC, and HIP (Finland); CEA and CNRS/IN2P3 (France); BMBF, DFG, and HGF (Germany); GSRT (Greece); OTKA and NIH (Hungary); DAE and DST (India); IPM (Iran); SFI (Ireland); INFN (Italy); MSIP and NRF (Republic of Korea); LAS (Lithuania); MOE and UM (Malaysia); BUAP, CINVESTAV, CONACYT, LNS, SEP, and UASLP-FAI (Mexico); MBIE (New Zealand); PAEC (Pakistan); MSHE and NSC (Poland); FCT (Portugal); JINR (Dubna); MON, RosAtom, RAS, RFBR and RAEP (Russia); MESTD (Serbia); SEIDI, CPAN, PCTI and FEDER (Spain); Swiss Funding Agencies (Switzerland); MST (Taipei); ThEPCenter, IPST, STAR, and NSTDA (Thailand); TUBITAK and TAEK (Turkey); NASU and SFFR (Ukraine); STFC (United Kingdom); DOE and NSF (USA).

\end{acknowledgments}

\clearpage
\bibliography{auto_generated}
\cleardoublepage \appendix\section{The CMS Collaboration \label{app:collab}}\begin{sloppypar}\hyphenpenalty=5000\widowpenalty=500\clubpenalty=5000\textbf{Yerevan Physics Institute,  Yerevan,  Armenia}\\*[0pt]
A.M.~Sirunyan, A.~Tumasyan
\vskip\cmsinstskip
\textbf{Institut f\"{u}r Hochenergiephysik,  Wien,  Austria}\\*[0pt]
W.~Adam, F.~Ambrogi, E.~Asilar, T.~Bergauer, J.~Brandstetter, E.~Brondolin, M.~Dragicevic, J.~Er\"{o}, M.~Flechl, M.~Friedl, R.~Fr\"{u}hwirth\cmsAuthorMark{1}, V.M.~Ghete, J.~Grossmann, N.~H\"{o}rmann, J.~Hrubec, M.~Jeitler\cmsAuthorMark{1}, A.~K\"{o}nig, I.~Kr\"{a}tschmer, D.~Liko, T.~Madlener, I.~Mikulec, E.~Pree, D.~Rabady, N.~Rad, H.~Rohringer, J.~Schieck\cmsAuthorMark{1}, R.~Sch\"{o}fbeck, M.~Spanring, D.~Spitzbart, J.~Strauss, W.~Waltenberger, J.~Wittmann, C.-E.~Wulz\cmsAuthorMark{1}, M.~Zarucki
% ...
\vskip\cmsinstskip
% ...
1:~~Also at Vienna University of Technology, Vienna, Austria\\
% ...

\end{sloppypar}
\end{document}